# PET GÖRÜNTÜLEME SİSTEMLERİNDE KULLANILAN TDC MİMARİLERİNİN VLSI BENZETİMLERİ

**Arş. Gör. Mehmet Akif ÖZDEMİR**
İzmir Kâtip Çelebi Üniversitesi, Mühendislik ve Mimarlık Fakültesi, Biyomedikal Mühendisliği
makif.ozdemir@ikc.edu.tr

**Prof. Dr. Ali TANGEL**
Kocaeli Üniversitesi, Mühendislik Fakültesi, Elektronik ve Haberleşme Mühendisliği
atangel@kocaeli.edu.tr

**Özet**

Pozitron emisyon tomografisi (PET), canlı bir vücut içindeki pozitron yayan radyonüklidlerin konsantrasyonlarının ölçülmesine dayanan tıbbi bir görüntüleme yöntemidir. PET görüntüleme sisteminde glikoz, pozitron yayan bir radyonüklid ile işaretlenip hastaya intravenöz olarak enjekte edilir. Pozitronlar doku içerisinde ilerleyip etkileştiği hücrelerin elektronları ile çarpışırlar. Bu etkileşim sonucu birbirine zıt yönde yayılan iki adet gama ışını oluşur. Radyoaktif glikozu tutmuş olan kanserli dokudan yayılan ışınlar halka şeklinde sıralanmış dedektörler aracılığı ile tespit edilir ve tespit edilen sinyaller elektriksel bir tepkiye dönüştürülür. Sonrasında bu tepkiler elektronik devreler ile örneklenir ve görüntü kümesi oluşturmak için histogram matrisleri şeklinde kaydedilir. Gama ışınları karşıt konumlandırılmış dedektörlere eşit sürede ulaşmıyor olabilir. Uçuş zamanı (TOF) özelliğine sahip olan PET'lerde tespit edilen ışınlar, iki fotonun dedektörlere ulaşma zamanı arasındaki farkın ölçülmesi prensibine dayanan bir yöntem ile daha iyi bir konumlama bilgisi elde etmeyi hedefler. Uçuş zamanının ölçülmesi işlemi zaman-sayısal dönüştürücü (TDC) yapıları ile gerçekleştirilir. Pikosaniye mertebesindeki bu zaman farkının ölçülme kabiliyeti PET sisteminin uzaysal çözünürlüğü ile doğrudan ilgilidir. Bu çalışmada PET sistemlerinde kullanım için çeşitli mimari yaklaşımına sahip TDC yapılarının 45 nm CMOS VLSI benzetimleri gerçekleştirilmiştir. Tasarlanan TDC mimarileriyle iki gama fotonunun dedektörlere ulaşma zamanı simülasyonu gerçekleştirilmiş ve zaman farkı başarılı bir şekilde sayısallaştırılmıştır. Ayrıca TDC mimarilerinin giriş çıkış gerilimleri, zamansal çözünürlükleri, ölçüm aralıkları ve güç analizleri gibi çeşitli performans ölçütleri belirlenmiştir. Tasarımı yapılan Vernier osilatör tabanlı TDC mimarisi ile 1V besleme geriliminin ve 1,62681 mW güç tüketiminde 25 ps zamansal çözünürlüklere ulaşılmıştır.

**Anahtar Kelimeler:** 45 nm, CMOS, Pozitron Emisyon Tomografisi (PET), Uçuş Zamanı (TOF), Vernier Osilatör, Zaman-Sayısal Dönüştürücü (TDC).

# VLSI IMPLEMENTATION OF TDC ARCHITECTURES USED IN PET IMAGING SYSTEMS

**Abstract**

Positron emission tomography (PET) is a medical imaging method based on the measurement of concentrations of positron emitting radionuclides in a living body. In the PET imaging system, glucose is labeled with a positron emitting radionuclide and injected intravenously. Then, the positrons move through the tissue and collide with the electrons of the cells in which they interact. As a result of this interaction, two gamma rays are emitted in the opposite direction. Gama rays emitted from cancerous tissue that have retained radioactive glucose are detected through ring-shaped detectors. And the detected signals are converted into an electrical response. Subsequently, these responses are sampled with electronic circuits and recorded as histogram matrix to generate the image set. The gamma rays may not reach the detectors located in the opposite position in equal time. In PETs having time of flight (TOF) characteristics, it is aimed to obtain better positioning information by a method based on the principle of measuring the difference between the reach time of the two photons to detectors. The measurement of the flight time is carried out with time-to-digital converter (TDC) structures. The measurement of this time difference at the ps level is directly related to the spatial resolution of the PET system. In this study, 45 nm CMOS VLSI simulations of TDC structures which have various architectural approaches were performed for using in PET systems. With the designed TDC architectures, two gamma photons time reach to detectors have been simulated and the time difference has been successfully digitized. In addition, various performance metrics such as input and output voltages, time resolutions, measurement ranges and power analysis of TDC architectures have been determined. Proposed Vernier oscillator-based TDC architecture has been reached 25 ps time resolution with a low power consumption of 1.62681 mW at 1V supply voltage.

**Keywords:** 45 nm, CMOS, Positron Emission Tomography (PET), Time of Flight (TOF), Vernier Oscillator, Time-to-Digital Converter (TDC).

## 1. GİRİŞ

İki fiziksel olay arasındaki zamanın kesin olarak elektronik ölçümü, birçok deneysel ve uygulamalı sistemde önemli ve esastır. TDC, iki rastgele olay arasındaki zaman farkını ölçebilen bir elektronik enstrümantasyon sistemidir(An, Son, An, & Kang, 2019; M. Lee, Kim, Park, & Sim, 2019; Sudo & Haneda, 2019; Wang, Dai, & Wang, 2018). Sayısal zaman dönüştürücüler, çok sayıda zaman ölçüm sistemlerinde yer edinmiştir ve endüstriyel, medikal, araştırma gibi çeşitli uygulama alanlarına sahiptir (Mhiri, Saad, Hammadi, & Besbes, 2017). Mesafe ölçümlerinin gerekli olduğu lazer radar sistemleri (Jahromi, Jansson, Nissinen, Nissinen, & Kostamovaara, 2015), analog sayısal çeviriciler (ADC) (Straayer & Perrott, 2008), kütle spektrometresi (Chernushevich, Loboda, & Thomson, 2001), manyetik rezonans görüntüleme sistemleri (Braga et al., 2013) ve PET tıbbi görüntüleme sistemleri (Cheng, Deen, & Peng, 2016; Kim et al., 2017) gibi birçok karışık sinyal devre uygulaması alanında TDC mimarisi yer bulur. Bu çalışmada açıklanan TDC mimarileri, PET tıbbi görüntüleme sistemlerinde kullanım içindir.

Bir PET tıbbi görüntüleme sistemi, insan vücuduna enjekte edilen radyonüklid madde sayesinde oluşan yüksek enerjili ışınların halka şeklinde sıralanmış algılayıcılar ile tespiti prensibi temeline dayanır. Algılayıcılarda tespit edilen yüksek enerjili ışınlar sayesinde ışımanın gerçekleştiği vücut dokusunun konumu tespit edilebilir ve pozisyonlanabilir. Kanserli bir dokuda meydana gelecek ışımaların tespiti ile bu dokunun sağlıklı dokulardan ayırt edilebilmesi sağlanabilir. Bir pozitron vücutta bir elektronla temas ettiğinde, iki parçacık yok olur ve 180 derece aralıklarla yayılan iki adet 511 KeV enerjili gama ışını üretir (Gambhir, 2002). Bir PET görüntüsü, bir pozitron ve elektronun yok edilmesinden ortaya çıkan kolineer 511 KeV enerjili gama ışını fotonlarının tespiti ile oluşturulur (Bailey, Maisey, Townsend, & Valk, 2005). PET tarayıcısında, birçok gama ışını dedektörü, Şekil 1'de gösterildiği gibi, görüntülenecek hastayı çevreleyen halkalar halinde konumlandırılmıştır.

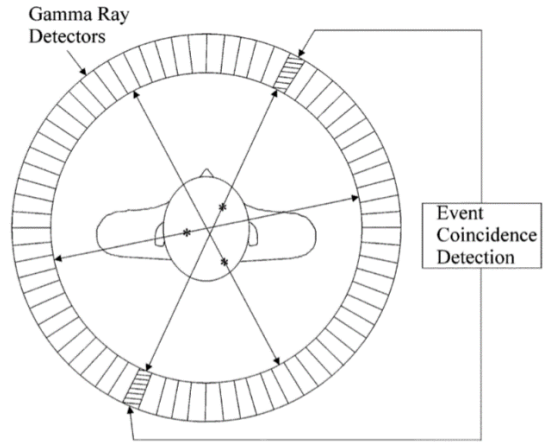

**Şekil 1**. Karşıt gama ışınlarının tespitinin gösterimi (Swann et al., 2004)

Pozitron elektron yok edilmelerinden oluşan karşıt gama ışınları bu dedektörler aracılığı ile algılanır ve algılanan sinyaller elektriksel bir tepkiye dönüştürülür. Bir PET tarayıcısından elde edilen görüntülerde, normal olmayan dokunun sağlıklı dokudan ayırt edilebilme kabiliyeti PET tarayıcısının görüntüleme kabiliyetini belirler. TOF PET sistemi yüksek enerjili ışınların karşılıklı sıralanmış halka şeklindeki algılayıcılara ulaşma süreleri arasındaki farktan daha iyi bir konumlama bilgisi çıkartmayı hedefler. Pozitron elektron yok edilmesi olayı karşılıklı sıralanmış halka dedektörlerine eşit mesafede gerçekleşmiyor olabilir. Bu nedenle yok olma olayı sonucu ortaya çıkan gama fotonları karşılıklı dedektörlere farklı zaman aralığında ulaşır. Yüksek enerjili ışınların algılayıcılara ulaşma süreleri arasındaki farklar pikosaniyeler mertebesindedir. TOF PET sisteminde pozisyonlama işlemi geleneksel PET sistemine göre daha hassas yapılmaktadır (Khalil, 2016). Geleneksel PET ve TOF PET sistemine ait uyarılma durumu Şekil 2'de gösterilmiştir. TOF PET sistemindeki zaman farkının hassas elektronik ölçümleri TDC'ler aracılığı ile yapılır. TOF PET sisteminde kullanılan TDC yapılarının zaman farkı ölçüm kabiliyetleri doğrudan görüntü çözünürlüğüne etki etmektedir.

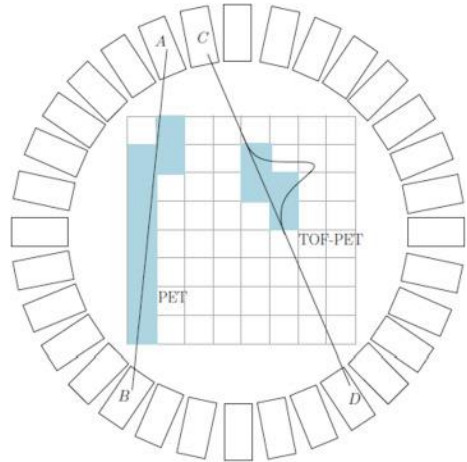

**Şekil 2.** Geleneksel PET ve TOF PET'de uyarılma durumu (Sitek, 2012)

Geleneksel PET'de, sistem zamanlama çözünürlüğü ($\Delta t$) bilgisi görüntü oluşturma sırasında pozisyonlama için kullanılmaz. TOF PET'de ise yok olma noktasının yeri yaklaşık olarak iki fotonun varış zamanlarındaki ($t_2 - t_1$) farkla belirlenir.

Gama fotonlarının dedektörlere ulaşma zamanları arasındaki fark Denklem (1)'de;

$$\Delta t = t_2 - t_1 \tag{1}$$

şeklinde ifade edilirse ve gama fotonlarının dedektörlere olan mesafe farkı ($\Delta d$);

$$\Delta d = c \frac{\Delta t}{2} \tag{2}$$

eşitliği ile tanımlanabilir. Denklem (2)'te gama fotonlarının 1 saniyede aldıkları yol ışık hızıdır ($3 \times 10^8$ cm/sn). Denklem (2)'ye göre 1 cm uzaysal çözünürlüğe ulaşmak için 66 ps zamansal çözünürlük gereklidir. TOF PET'te belli zaman penceresinde gelen yok olma olayı sonucu oluşan gama fotonlarının, karşılıklı detektörlere ulaşma zaman farkı ölçümleri TDC'ler aracılığı ile yapılabilmektedir (Rolo et al., 2013). Bu çalışmada PET tıbbi görüntüleme sistemlerinde kullanım için TDC mimarilerinin çok geniş ölçekli tümleşim (Very Large Scale Integration, VLSI) benzetimleri gerçekleştirilmiştir. Tasarlanan TDC mimarileri ile PET görüntülerine daha iyi çözünürlük sağlayarak kanserli dokuların sağlıklı dokulardan daha belirgin bir şekilde ayırt edilebilmesi amaçlanmaktadır.

## 2. MATERYAL VE METOD

PET, halka şeklinde dizilmiş bir seri sintilasyon kristali ve bu kristallere birleştirilmiş foton çoğaltıcı tümlerden (PMT) oluşmuştur. PMT'den çıkan sinyali yükseltmek için kullanılan sinyal kuvvetlendirici devreler ve anhilasyon darbelerinin yakalanması için kullanılan sinyal yükseklik analizörleri (SCA), PET sisteminin temel bileşenlerindendir. Ayrıca pozisyonlama için TDC-FPGA sisteminden oluşan sayısallaştırma devreleri ve nihai görüntüyü elde etmek ve düzenlemek için kullanılan bilgisayar ünitesi ve görüntüleme sistemi PET sistemlerinde kullanılmaktadır (B. J. Lee, Chang, & Levin, 2018). Şekil 3'de PET cihazında anhilasyon olayının tespitinden, görüntü oluşumuna kadar olan aşamalar verilmiştir.

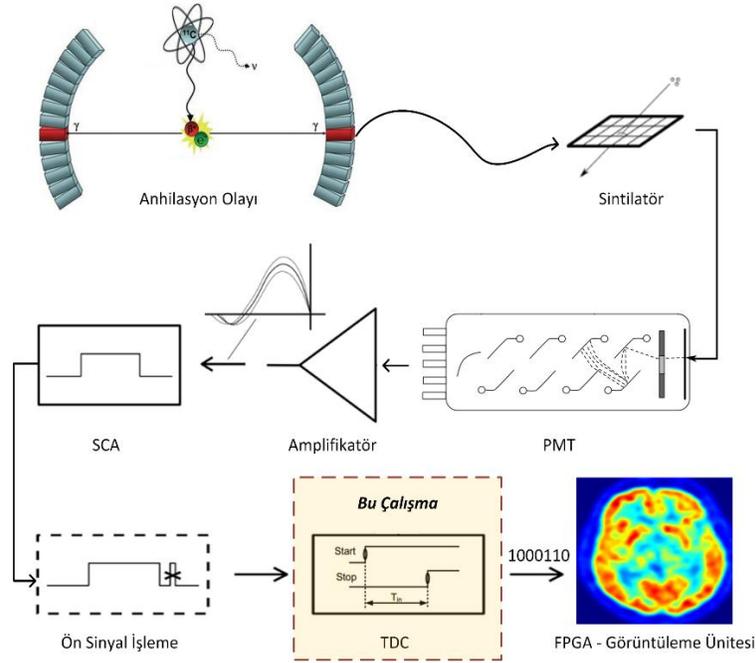

**Şekil 3.** PET sistemi

### 2.1. TDC Mimarisi

TDC, iki sinyal arasındaki küçük zaman farklarını ("Başlat" ve "Durdur" olarak tanımlanır) ölçen ve bu zaman aralığının sayısal gösterimlerini sağlayan bir elektronik entegre devredir. TDC'nin temel blok diyagramı Şekil 4(a)'da gösterilmiştir. Bir TDC'nin işlevi ADC ile aynıdır. TDC, ADC'lerdeki voltaj veya akım farkları yerine zaman farkıyla ilgilidir. Şekil 4(b)'de görüldüğü gibi ölçülen süre, Başlat ve Durdur'un pozitif kenarları arasındaki faz farkı olarak tanımlanır (Zanuso, Levantino, Samori, & Lacaita, 2010). Giriş sürekli zaman sinyalleridir. Çıkışlar sayısal kodlardır.

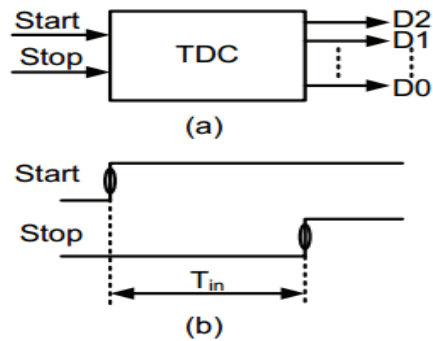

**Şekil 4.** Zaman-sayısal dönüşümünün temeli (Gao, Gao, Hu-Guo, & Hu, 2011)

Bu çalışmada Vernier gecikme hattı (VDL) tabanlı bir TDC mimarisi yaklaşımı önerilmektedir. Tasarımı yapılan VDL TDC mimarisinin şematik gösterimi Şekil 5'de gösterilmiştir. Bu mimaride ölçümler Vernier cetveli ölçüm prensibine dayanmaktadır. Farklı evirici yapılarından oluşturulan gecikme süreleri farklı iki gecikme hücresi, iki farklı gecikme hattına yerleştirilir. Vernier yöntemi kullanılarak iki gecikme hücresinin gecikme süreleri arasındaki fark ölçülebilir. Geciken sinyallerin hangisinin devreye daha erken geldiği karar verici devreler aracılığı ile kaydedilir ve örneklenir (Wang & Dai, 2017).

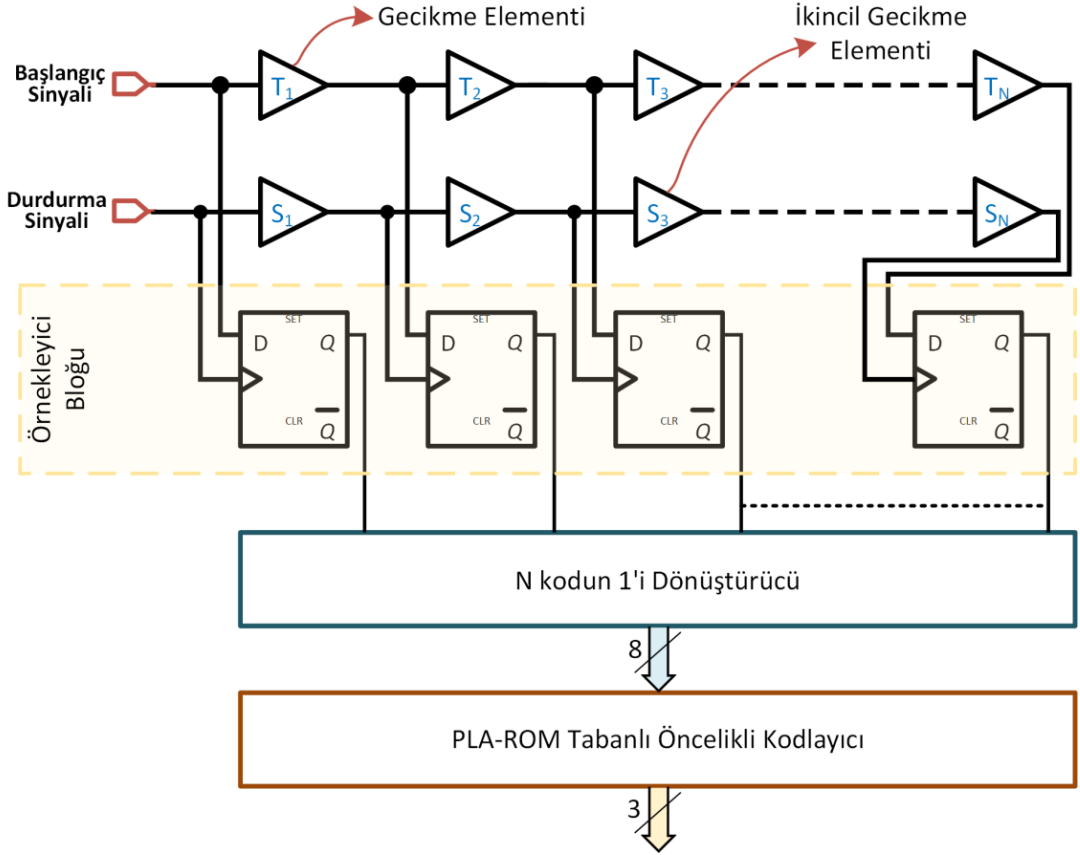

**Şekil 5.** Vernier gecikme hattı tabanlı TDC blok diyagramı

Bu mimaride başlangıç sinyali ve durdurma sinyali yok olma olayı sonucu oluşan iki gama fotonunun karşılıklı sıralanmış dedektörlere ulaşma zamanını temsil etmektedir. Başlangıç sinyali gecikme elementi olarak nitelendirilen katlı evirici yapısından geçirilerek geciktirilir. Eş zamanlı olarak durdurma sinyali de ikincil gecikme elementi olarak nitelendirilen evirici yapısından geçirilerek geciktirilir. Başlangıç sinyali durdurma sinyaline göre daha fazla evirici yapısından geçirildiği için daha çok gecikmeye maruz kalır. Geciktirilen sinyallerin birbirine göre konumlarından o anki zaman farkı ölçümü sayısallaştırılabilir. Örnekleyici bloğu temel D flip flop yapılarından oluşmaktadır. Geciken başlangıç sinyalleri geciken durdurma sinyallerinden önce örnekleyici bloğuna ulaşırsa örnekleyici çıkışları lojik-1 seviyesine ayarlanır. Geciken başlangıç sinyalleri geciken durdurma sinyallerinden daha sonra örnekleyici bloğuna ulaşırsa örnekleyici çıkışları lojik-0 seviyesine ayarlanır. İki sinyal arasındaki mesafe bazlı bu ölçüm Vernier prensibi olarak tanımlanır. Örnekleyici bloğunun çıkışları termometre kod dizişi şeklindedir. Bu çıkışlar sonrasında N kodun 1'i dönüştürücü ve öncelikli kodlayıcı yapısıyla termometre kodundan sayısal ikilik kodlara dönüştürülür.

## 3. SONUÇ

Tasarlanan VDL TDC mimarisinin 45nm CMOS transient simülasyon analizi Şekil 6'da gösterilmiştir. 'T' harfiyle nitelendirilen gecikmiş başlangıç sinyalleri 'S' harfiyle nitelendirilen gecikmiş durdurma sinyallerinden ilk 5 gecikmede örnekleyici bloğuna daha önce ulaşmıştır. Bu sinyaller dijital çıkışlara lojik-1 olarak atanmıştır. Kalan yüksek seviyeli sinyaller ise lojik-0 olarak atanmıştır. Öncelikli kodlayıcı yapısı ise bu sinyalleri '101' şeklinde ikilik çıkış olarak sayısallaştırmıştır.

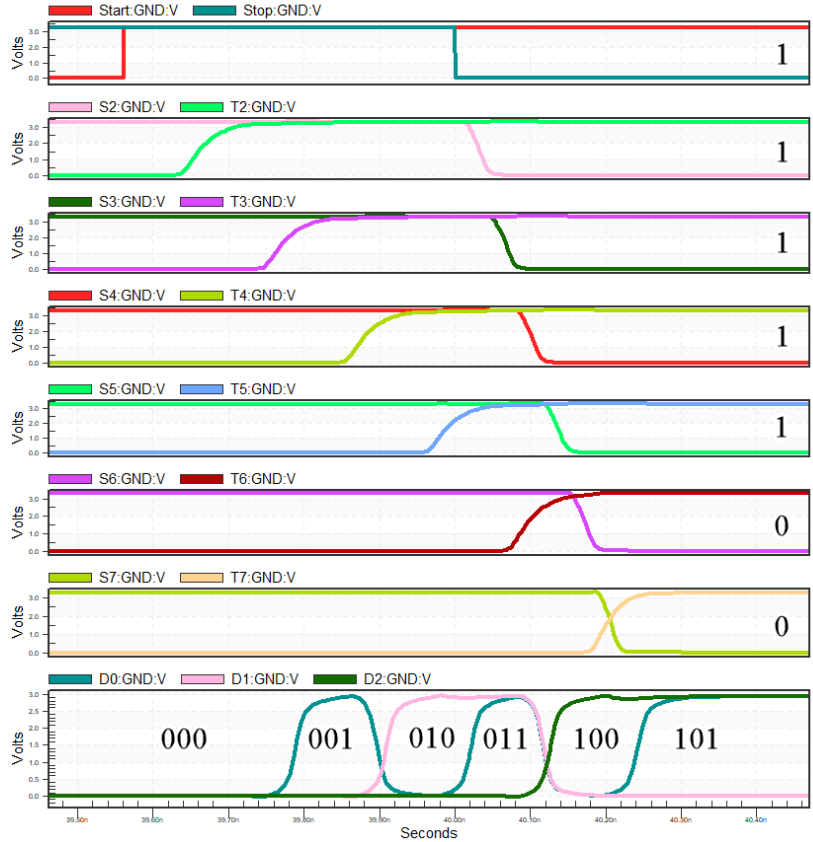

**Şekil 6.** Vernier TDC transient simülasyon analizi

## 4. TARTIŞMA

Bu çalışmada PET tıbbi görüntüleme cihazlarında kullanım için 45nm CMOS teknoloji boyutunda VDL tabanlı TDC mimarisinin tasarımı gerçekleştirilmiş ve tasarlanan mimarinin giriş çıkış gerilim ilişkisi ile güç analizleri simülasyonları gerçekleştirilmiştir. VDL TDC mimarisinde TDC çözünürlüğü iki gecikme elementinin gecikme sürelerine bağlıdır. Tasarımı yapılan mimaride başlangıç sinyali 2,5 ns anından başlayarak ~102,7 ps aralıklarla geciktirilmiştir. Durdurma sinyali ise 4 ns anından başlayarak ~77,7 ps zaman aralıklarıyla geciktirilmiştir. Bu sayede tasarlanan mimarinin zamansal çözünürlüğü 102,7-77,7=~25 ps mertebesinde gerçekleşmiştir. Tasarımda uzunluğu 45 nm genişliği 1,5 um 429 adet özdeş MOSFET kullanışmış olup 1 V besleme geriliminde ortalama güç tüketimi çok düşük seviyede, 1,62681 mW olarak geçekleşmiştir. Önerilen VDL tabanlı TDC mimarisi ulaştığı 25 ps zamansal çözünürlük ve çok düşük güç tüketimi ile PET görüntüleme sistemlerinde kullanım için tasarlanacak tam bir çip yapısına kaynak oluşturabilir. Bununla birlikte TDC'nin zamansal çözünürlüğü evirici tabanlı gecikme elementlerinin Gate gecikmesine bağlı olduğu için tasarımı yapılacak daha düşük Gate gecikmesine sahip yapılarla ile daha hassas zamansal çözünürlüklere erişilebilir.

# KAYNAKLAR